\documentclass[prl,aps,showpacs,twocolumn,floatfix]{revtex4}

\usepackage{graphicx}
\usepackage[ansinew]{inputenc}
\usepackage{array}
\usepackage{color}
\usepackage{amsmath}
\usepackage{amsxtra}
\usepackage{amstext}
\usepackage{amssymb}
\usepackage{latexsym}
\usepackage{dsfont}
\begin{document}

\title{Using a Laguerre-Gaussian beam to trap and cool the rotational motion of a mirror}

\author{M. Bhattacharya and P. Meystre}
\affiliation{B2 Institute, Department of Physics and College of Optical
Sciences\\The University of Arizona, Tucson, Arizona 85721}

\date{\today}

\begin{abstract}
We show theoretically that it is possible to trap and cool the
rotational motion of a macroscopic mirror made of a perfectly 
reflecting spiral phase element using orbital angular momentum 
transfer from a Laguerre-Gaussian optical field. This technique 
offers a promising route to the placement of the rotor in its 
quantum mechanical ground state in the presence of thermal noise. 
It also opens up the possibility of simultaneously cooling a 
vibrational mode of the same mirror. Lastly, the proposed 
design may serve as a sensitive torsional balance in the quantum 
regime.
\end{abstract}


\maketitle

The optical control of the quantum mechanical center-of-mass
motion of ions, neutral atoms, molecules and microscopic-scale objects has
formed a dominant theme in atomic, molecular and optical physics
in recent years \cite{letokhovbook,meystrebook}. These techniques
are now being transferred to the macroscopic regime, where laser
light has been found to be effective in cooling and trapping, for
example, gram-scale mirrors \cite{corbitt2007}. Such efforts
are part of the rapidly emerging field of `quantum optomechanics',
one of the ambitious aims of which is to place a macroscopic
object in its quantum mechanical ground state
\cite{gigan2006,kleckner2006,arcizet2006,schliesser2006}. The
accomplishment of this task would open up a host of fascinating
scenarios ranging from the fundamental
\cite{leggett2002,marshall2003, hansch2007} to the applied
\cite{caves1980,courty2003,walls1994}, all following from the
manifestation of quantum mechanical behavior in classical objects.

So far most efforts have been aimed at quantizing linearly
vibrating oscillators, which are prototypical mechanical objects
\cite{gigan2006,kleckner2006,arcizet2006}. Further, oscillators
with breathing modes have been considered, driven by the
centrifugal effects of radiation pressure
\cite{carmon2005,schliesser2006,carmon2007}. A torsional mode has
also been examined experimentally \cite{tittonen1999}, but its
effective dynamics was essentially vibrational in that the
interaction mechanism was the mirror's exchange of \textit{linear}
momentum with radiation, a fact true of all the examples
cited above.

In this Letter we point out the possibility of quantizing instead
a {\em rotational} mode of a classical torsional oscillator using
the exchange of \textit{angular} momentum with a radiation 
field. Combined with the use of the linear momentum of light, this 
results in the potential to simultaneously quantize a vibrational as 
well as a rotational mode of motion of an oscillator using the same 
radiation field. Also, (torque-induced) rotation of the mirror 
manifests itself as a change in cavity length, which may be
measurable to a sensitivity better than the quantum noise limit, 
based on recent analyses \cite{chen2002,pinard2006}. We therefore 
expect our proposed design to possibly find use as a sensitive 
torque sensor, (i.e. a torsion balance) operating in the quantum 
regime.

Specifically we consider the mechanical effect of a 
Laguerre-Gaussian beam interacting with a macroscopic mirror. 
In addition to their intrinsic spin angular momentum $\leq \hbar$, 
photons in such beams also carry an integral orbital angular momentum 
$l\hbar$, see \cite{OAMbook} and references therein. It has been 
experimentally demonstrated that Laguerre-Gaussian beams can exert a torque on 
microscopic particles \cite{he1995,volpe2006} as well as on a 
Bose-Einstein condensate, where they create vortices \cite{andersen2006}. 
The main result of our analysis is to demonstrate that the transfer 
of orbital angular momentum can be large enough to have a significant 
effect on macroscopic objects as well and can dramatically cool their 
rotational motion.

\begin{figure}[h!]
\includegraphics[width=0.48 \textwidth]{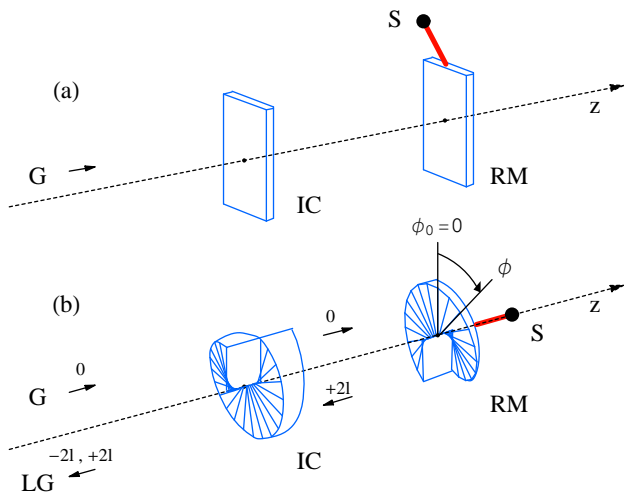}
\caption{\label{fig:cavitypic}(Color online).(a) A typical
arrangement for trapping and cooling a vibrating mirror using
linear momentum transfer from a Gaussian cavity mode (G). The
cavity is formed by a fixed partially transparent input coupler 
(IC) and a movable perfectly reflecting rear mirror (RM), the 
latter being suspended from a support S. The RM vibrates along the 
$z-$axis. (b) The arrangement proposed in this work for trapping 
and cooling a rotating mirror using angular momentum exchange with 
a Laguerre-Gaussian (LG) mode. The cavity is formed by the IC and 
the RM, both spiral phase elements. The RM is mounted on a
support S and can rotate about the $z-$axis. Its angular 
deflection from equilibrium $(\phi_{0}=0)$ is indicated by the 
angle $\phi$. The charge on the beams at various points has 
been indicated.}
\end{figure}

The typical experimental set-up used to achieve radiation cooling
and trapping of a \textit{vibrating} mirror \cite{gigan2006} is
schematically shown in Fig.\ref{fig:cavitypic}(a).
It consists of an optical cavity where the input coupler 
is fixed and the rear mirror is suspended from a support 
and allowed to oscillate harmonically along the cavity axis. Usually 
the input coupler is weakly transmissive and the rear mirror is taken 
to be perfectly reflecting. Also, the input coupler is massive and 
the rear mirror micro- or nano-fabricated. The rear mirror can then be 
trapped (cooled) using a Gaussian laser beam tuned above (below) a 
cavity resonance.

We consider the rotational analog of Fig.\ref{fig:cavitypic}(a) 
shown in Fig.\ref{fig:cavitypic}(b). The optical mirrors of 
the vibrating cavity have been replaced by spiral phase 
elements which are commonly used to modify the azimuthal 
structure of laser beams either via reflection or transmission 
\cite{Hooft2004}. Given the wavelength of radiation being used they 
can be designed to impart a fixed topological charge to an incident beam. 
In what follows we assume a Gaussian beam, which has charge 0, to be
incident on the cavity. The azimuthal structure of the input coupler is such 
that upon reflection from either side it \textit{removes} a fixed charge  
$2l$ from the beam. The input coupler is also partially transparent and since 
the two sides have opposite winding it allows beams from either 
side to pass through with no change to their charge -- an effect  
that has been experimentally observed for a Gaussian beam \cite{Hooft2004}. 
The perfectly reflecting rear mirror on the other hand, is designed so that 
it \textit{adds} a charge $2l$ to a beam. With these specifications 
in mind we follow the input beam through the set-up of 
Fig.\ref{fig:cavitypic}(b) where the charges at various positions 
along the $z-$axis have been indicated.

The Gaussian input beam first meets the input coupler. The reflected component has a 
charge $-2l$ while the transmitted beam has charge 0. The charge 0 
beam reflected from the rear mirror gets charged to $2l$. Returning to the input coupler, 
reflection gives a mode with charge 0, while transmission gives a beam 
with charge $2l$. That this configuration provides self-consistent 
conditions for intra-cavity mode build-up has been noted before 
\cite{LaserResonators3book}.

We assume that the rear mirror is a thin disk of mass $M$ and radius $R$,
so that its moment of inertia about the $z-$axis passing through
its center is given by $I=MR^{2}/2$. The rear mirror is part of a
torsional pendulum mounted on a support S and oscillating with angular
frequency $\omega_{\rm \phi}$. It has an equilibrium position
$\phi_{\rm 0}=0$ in the absence of radiation, and undergoes angular
deviations $\phi$ from that position that are small enough that
they can be described harmonically $(\phi \ll 2\pi)$.

Coupling between radiation and the rear mirror occurs because at each 
reflection a charge 0 photon acquires angular momentum $2l\hbar$ 
from the mirror. This happens once in every cavity round trip time 
$2L/c$, where $L$ is the length of the cavity and $c$ is the 
velocity of light. The radiation torque on the rear mirror equals the 
angular momentum change per unit time 
$2l\hbar/(2L/c)=\hbar\xi_{\rm \phi}$, where we refer to
$\xi_{\rm \phi}=cl/L$ as the \textit{optorotational} coupling
parameter. Accounting for the intracavity photon number the 
coupling between the radiation field and the mirror can be 
included in the model Hamiltonian
\begin{equation}
\label{eq:Ham2}
H_{\rm \phi} = \hbar \omega_{c}a^{\dagger}a+\frac{L_{z}^{2}}{2I}
+\frac{1}{2}I\omega_{\rm \phi}^{2}\phi^{2}
-\hbar\xi_{\rm \phi} a^{\dagger}a\phi.
\end{equation}
Here $a$ ($a^\dagger$) are the bosonic annihilation (creation) 
operators for the cavity mode of frequency $\omega_{c}$ satisfying 
the commutation relation $[a,a^\dagger]=1$. $L_{z}$ is the angular 
momentum of the rear mirror about the intra-cavity axis and
satisfies the commutation relation $[L_{z},\phi]=-i\hbar$. 
The first term in $H_{\phi}$ describes the energy of the cavity mode, 
the next two terms the energy of the oscillating rear mirror and the last 
term the effect of radiation torque on the rear mirror.

The evolution of the dynamical variables in the
Hamiltonian~(\ref{eq:Ham2}) is given by the Heisenberg equations
of motion, to which we add damping and noise and obtain the
corresponding quantum Langevin equations in the standard way
\cite{gardinerbook}:
\begin{eqnarray}
\label{eq:QLErot}
 \begin{array}{ll}
\dot{a}= &-i(\delta-\xi_{\phi}\phi)a-\frac{\gamma}{2}a+\sqrt{\gamma}a^{\rm in},\\
\dot{\phi}= & L_{z}/I,\\
\dot{L_{z}}= & -I\omega_{\phi}^{2}\phi+\hbar \xi_{\phi}a^{\dagger}a
-\frac{D_{\phi}}{I}L_{z}+\epsilon_{\phi}^{\rm in}.\\
\end{array}
\end{eqnarray}
Here $\delta =\omega_{c}-\omega_{L}$ is the detuning of the laser
frequency $\omega_{L}$ from the cavity resonance, $\gamma$ is the
damping rate of the cavity, $D_{\phi}$ the intrinsic
damping constant of the oscillating rear mirror and $a^{\rm in}$ 
is a noise operator describing the laser field incident on the cavity. 
The mean value $\langle a^{\rm in}(t)\rangle =a_{\rm s}^{\rm in}$
describes the classical Gaussian field, and the delta-correlated
fluctuations $\langle \delta a^{\rm in}(t) \delta a^{\rm in,\dagger}(t')
\rangle=\delta (t-t')$ add vacuum noise to the cavity modes. The 
Brownian noise operator $\epsilon_{\phi}^{\rm in}$ represents the 
mechanical noise that couples to the mirror from the environment. 
Its mean value is zero and its fluctuations are correlated at 
temperature $T$ as
\cite{gardinerbook}
\begin{eqnarray}
\label{eq:Brownian}
\begin{array}{l}
\langle \delta \epsilon_{\phi}^{\rm in}(t) \delta \epsilon_{\phi}^{\rm in}(t') \rangle=\\
D_{\phi}\int_{-\infty}^{\infty} \frac{d\omega}{2\pi}e^{-i\omega(t-t')}\hbar \omega
\left[1+ \coth \left(\frac{\hbar \omega}{2k_{\rm B}T}\right)\right],\\
\end{array}
\end{eqnarray}
where $k_{\rm B}$ is Boltzmann's constant. The steady-state 
values of the dynamical variables are easily found to be
\begin{equation}
\label{eq:sstate}
a_{\rm s}=\frac{\sqrt{\gamma}|a_{\rm s}^{\rm in}|}{\left[(\frac{\gamma}{2})^{2}+(\delta-\xi_{\phi}\phi_{\rm s})^{2} \right]^{1/2}},
\,\,\,\phi_{\rm s}=\frac{\hbar \xi_{\phi}a_{\rm s}^{2}}{I\omega_{\phi}^{2}}, \,\,\,L_{z,\rm s}=0,\\
\end{equation}
where the phase of the input field $a_{\rm s}^{\rm in}$ has been
chosen such that $a_{\rm s}$ is real. For future use we define the
input power as $P_{\rm in}=\hbar \omega_{c}|a_{\rm s}^{\rm in}|^{2}$.
Note that the coupled steady-state equations for $a_{\rm s}$ and
$\phi_{\rm s}$ can be solved explicitly to yield solutions that
display bistability for high enough $P_{\rm in}$
\cite{pm1985,mancini1994}. In the following we will use static
feedback to maintain the length of the cavity and hence remove 
bistability.

To examine the displacement of the system from steady-state under
the influence of noise we expand every operator as the sum of a
mean value corresponding to its steady state
[Eq.(\ref{eq:sstate})] and a small fluctuation. For example,
$a=a_{\rm s}+\delta a$. Inserting analogous expressions for all
operators into Eq.~(\ref{eq:QLErot}) and linearizing the resulting
equations in the fluctuations yields the set of equations
\begin{equation}
\label{eq:fluctrot} \dot{u}(t)=B u(t)+n(t),
\end{equation}
where the vector of fluctuations is $u(t)=(\delta X_{a},\delta
Y_{a},\delta \phi,\delta L_{z})$, and the input noise vector is
$n(t)=(\sqrt{\gamma}\delta X_{a}^{\rm in},\sqrt{\gamma}\delta
Y_{a}^{\rm in},0,\delta \epsilon_{\phi}^{\rm in})$, where we have
redefined the field fluctuations in terms of their quadratures as
$\delta X_{a} =(\delta a+\delta a^{\dagger})/\sqrt{2}$, $\delta
Y_{a} =(\delta a-\delta a^{\dagger})/i\sqrt{2}$, etc. 

Further, $B$ in Eq.(\ref{eq:fluctrot}) is a matrix (we do not reproduce it 
here for reasons of space) 
%\begin{equation}
%\label{eq:RHmatrix} B =
%\begin{pmatrix}
%-\gamma/2  &  \Delta             &  0                    &       0   \\
% -\Delta            & -\gamma/2  & \sqrt{2} a_{\rm s} \xi_{\phi}    &       0   \\
%0               &    0             &     0                      &  1/I\\
%\sqrt{2}\hbar a_{\rm s} \xi_{\phi} &    0     & -I D_{\phi}^{2} &-D_{\phi}/I\\
%\end{pmatrix},
%\end{equation}
whose elements have to satisfy a sequence of algebraic inequalities 
which arise from the conditions for dynamical stability of the steady 
state solutions [Eq.(\ref{eq:sstate})] formalized in terms of the 
Routh-Hurwitz criterion \cite{dejesus1987}. Below we will choose 
two laser fields, one for cooling and one for trapping, so that 
their combined effect stabilizes the mirror depending on the values 
of the effective moving mirror frequency and damping, which we now 
calculate.

We solve Eq.~(\ref{eq:fluctrot}) by taking its Fourier transform
and reducing it to a set of algebraic equations. Specifically we
solve for the fluctuations in the displacement
\begin{equation}
\label{eq:disp}
\delta \phi (\omega) = \chi (\omega)\delta \tau(\omega),
\end{equation}
where $\chi(\omega)$ is the susceptibility of the oscillating rear mirror and 
$\delta \tau(\omega)$ the total fluctuation in the torque acting on it, 
including both radiation and mechanical contributions \cite{cohadon1999}. 
The susceptibility has a Lorentzian shape,
$\chi^{-1}(\omega)=I(\omega_{\rm
eff}^{2}-\omega^{2})-iD_{\rm eff}\omega$, where the
effective frequency and damping are given by
\begin{eqnarray}
\label{eq:eff}
\begin{array}{ll}
\omega_{\rm eff}^{2}=&\omega_{\phi}^{2}-\frac{2\xi_{\phi}^{2} \gamma P_{\rm in}}{I\omega_{c}}
\frac{\Delta}{\Delta^{2}+\frac{\gamma^{2}}{4}}
\frac{(\frac{\gamma}{2})^{2}-(\omega^{2}-\Delta^{2})}{\left[(\frac{\gamma}{2})^{2}+(\omega-\Delta)^{2}\right]
\left[(\frac{\gamma}{2})^{2}+(\omega+\Delta)^{2}\right]},\\
D_{\rm eff}=&D_{\phi}+\frac{2\xi_{\phi}^{2} \gamma P_{\rm in}}{\omega_{c}}
\frac{\Delta}{\Delta^{2}+\frac{\gamma^{2}}{4}}
\frac{\gamma}{\left[(\frac{\gamma}{2})^{2}+(\omega-\Delta)^{2}\right]
\left[(\frac{\gamma}{2})^{2}+(\omega+\Delta)^{2}\right]},\\
\end{array}
\end{eqnarray}
where $\Delta=\delta-\xi_{\phi}\phi_{\rm s}$ is the full static
detuning. The first and second terms on the right hand side of each of the
Eqs.~(\ref{eq:eff}) represent the mechanical and optomechanical
contributions respectively to the effective frequency and damping
of the oscillating rear mirror. We note that the latter contributions 
are proportional to $\xi_{\phi}^{2} \propto l^{2}$. The values 
of $\omega_{\rm eff}$ and $\gamma_{\rm eff}$ we will use below 
are found by evaluating Eq.(\ref{eq:eff}) at 
$\omega=\omega_{\rm \phi}$.

As shown recently, $\omega_{\rm eff}$ and $\gamma_{\rm eff}$ can
be controlled almost independently by driving the cavity with two
laser fields of different powers and detunings \cite{corbitt2007}.
An intense `trapping' field detuned (relatively) far above the
cavity resonance $(\Delta <0)$ makes the spring stiffer while
introducing a slight anti-damping. A less intense `cooling' field
detuned below the cavity resonance $(\Delta >0)$ increases
damping, overcompensating for the trapping field at the cost of
also causing a small amount of anti-trapping. We propose a similar
configuration here, with the trapping and cooling laser fields
both being Gaussian beams.

To find the number of quanta of excitation of the rotor, we consider 
Eq.(\ref{eq:disp}). For our model the torque fluctuations 
are chiefly due to Brownian noise, i.e. 
$\delta \tau \sim \delta \epsilon_{\phi}^{\rm in}$. Further
we will consider mirror cooling at temperatures 
$T \gg \hbar \omega_{\rm eff}/k_{\rm B}$, 
so the high-temperature limit of Eq.(\ref{eq:Brownian}) may be used. 
In frequency-space this yields 
$\langle \delta \tau(\omega)\delta \tau(\omega')\rangle=
2D_{\phi}k_{\rm B}T \delta (\omega +\omega')$, which when 
combined with Eq.(\ref{eq:disp}) and Fourier-transformed gives
$\langle \delta \phi(t) \delta \phi(t') \rangle = \frac{D_{\phi}k_{\rm B}T}{\pi} 
\int_{-\infty}^{\infty} d\omega \,e^{-i\omega(t-t')} \,|\chi(\omega)|^{2}$ 
\cite{gardinerbook}. We set $t=t'$ in this expression and use the 
Lorentzian form of the susceptibility to find 
$\int_{-\infty}^{\infty} d\omega \,|\chi(\omega)|^{2} = 
\pi/(I\omega_{\rm eff}^{2}D_{\rm eff})$. We then introduce the
equipartition theorem to relate $\langle \delta \phi(t)^{2}\rangle $  
to $T_{\rm eff}$, the effective temperature of the rear mirror : 
$k_{\rm B}T_{\rm eff}/2 = I \omega^{2}_{\rm eff}\langle \delta \phi^{2}(t)\rangle/2.$
This implies
\begin{equation}
\label{eq:teff}
T_{\rm eff}=\left(\frac{D_{\rm \phi}}{D_{\rm eff}}\right)T,
\end{equation}
and therefore the number of rotational quanta of the oscillating rear mirror 
\cite{mbpm2007}
\begin{equation}
 \label{eq:quanta}
n=\frac{k_{B}T_{\rm eff}}{\hbar \omega_{\rm eff}}= \frac{k_{B} T}{\hbar \omega_{\rm eff}}
\left(\frac{D_{\phi}}{D_{\rm eff}}\right).
\end{equation}
We consider as an example a $10\mu$g mirror of radius $10\mu$m mounted
on a torsional support with angular frequency $\omega_{\phi}=2\pi
\times 2.5$kHz and a mechanical quality factor $\sim 10^{5}$. At 
room temperature $T=300$K, this rotor has $n \sim 3\times 10^{8}$ quanta. 

We make the moving mirror part of a cavity $L=1$mm long. We 
choose $l=100$ and a Gaussian beam with $P_{\rm in}^{t}=250$mW 
at a detuning $\Delta^{t}=-2.5\gamma$ to trap and another 
Gaussian field with $P_{\rm in}^{c}=4$mW at a detuning 
$\Delta^{c}=0.5\gamma$ to cool the moving mirror, where 
$\gamma=2\pi \times 10$MHz implies an optical cavity finesse 
of $\sim 10^{4}$. This yields 
$\omega_{\rm eff}\sim 10\omega_{\phi}$, $D_{\rm eff} \sim 4\times 10^{4} D_{\phi}$ 
and $T_{\rm eff}\sim 8$mK, indicating that the radiation can both 
trap and cool the mirror. The number of quanta is lowered to 
$n \sim 10^{2}$. In these calculations we have assumed static 
$(\omega=0)$ feedback to maintain the cavity length, which 
gives us control over $\Delta$ independent of input power 
as is usually implemented in the case of vibrational cooling 
\cite{corbitt2007}. With a mechanical quality factor twice as 
high and cryogenic lowering of the base temperature to 
$T \sim 3$K, $n<1$ can be achieved, i.e. the mirror can be 
brought to its ground state. 

Having demonstrated that the proposed technique is promising 
for achieving ground state occupation by the rotor, we make three
comparisons to the standard case of vibrational cooling using linear 
momentum. First, in practice it is not easy to increase the linear 
momentum of the cooling photons (it usually entails changing the laser), 
but high-$l$ Laguerre-Gaussian modes can be achieved readily \cite{OAMbook}. 
Further the optomechanical effects of rotational cooling enjoy a quadratic 
scaling with the angular momentum as shown above, while in the 
vibrational case the scaling goes linearly with the linear momentum
\cite{corbitt2007}. Thus a Laguerre-Gaussian mode with sufficiently 
large charge can bring the rotor to its ground state even 
with the use of moderate laser power. This is an advantage from the 
point of view of mirror heating effects.

Second, an interesting scenario arises if the rear mirror is allowed 
to vibrate in addition to rotating. Since it can possess linear as 
well as angular momentum, the same light field can then couple to both 
modes of mirror motion. 

Lastly, it has recently been predicted that the sensitivity of a detuned
cavity with a linearly vibrating mirror to a force causing a change in 
the cavity length can be improved beyond the standard quantum noise limit
\cite{chen2002,pinard2006}. This analysis can be adapted to our case 
and implies that the design presented here may be useful as a sensitive 
torque-to-displacement transducer in the limit where classical noise has 
been suppressed, i.e. in the quantum regime.

In conclusion we have demonstrated that orbital angular momentum
exchange with a Laguerre-Gaussian mode enables a classical rotor
to release its excitation quanta, in principle until it reaches
its ground state.
%\begin{acknowledgments}
This work is supported in part by the US Office of Naval Research,
by the National Science Foundation, and by the US Army Research
Office. We would like to thank H. Uys, O. Dutta and C. Maes for
stimulating discussions.

\textit{Note:} There appeared in Ref.~\cite{sun2007}, after the
submission of this work, a proposal for using electronic spin-orbit 
interactions in semiconductors to cool torsional nanomechanical 
vibrations. 
%\end{acknowledgments}

%\end{thebibliography}
\end{document}